\documentstyle [12pt]{article}
\topmargin 
.15in
\renewcommand{\baselinestretch}{1} 
\title{One-loop QCD Correction for
 Inclusive Jet Production and Drell-Yan Process  in Composite
Models}

\author{ Taekoon Lee 
        \\
        \\
        Fermi National Accelerator Laboratory\\
                P.O. Box 500, Batavia, IL 60510}

\date{}
\begin{document}
\maketitle
\begin{abstract}
We calculate the one-loop QCD 
correction for inclusive jet 
cross section and Drell-Yan process in a general low-energy 
effective Lagrangian for composite
quarks and leptons.

\end{abstract} 

\def\thepage{FERMILAB-PUB-96/117-T}
\thispagestyle{myheadings}
\newpage
\renewcommand{\baselinestretch}{2}
\pagenumbering{arabic}
\addtocounter{page}{0}
\newcommand{\be}{\begin{equation}}
\newcommand{\ee}{\end{equation}}
\newcommand{\bear}{\begin{eqnarray}}
\newcommand{\eear}{\end{eqnarray}}

\section{Introduction}

Recently the CDF group at Fermilab reported a significant excess 
of
one-jet inclusive production at high $p_{T}$ over the standard 
QCD
prediction \cite{r1}. The observed inclusive jet cross section 
is
in excellent agreement with theory, but  begins to
deviate from the QCD prediction around $p_{T}=200$ GeV,  and 
its central
value becomes as large as twice  the theoretical prediction 
at 
$p_{T}$ above $400$ GeV.  If this discrepancy between theory 
and 
experiment survives more stringent
tests, and arises not from the uncertainties
in the QCD parameters such as the parton distribution functions 
but
from a genuine new physics, one possible new physics explanation 
would
be  existence of a quark substructure.  

A substructure in quarks gives rise to 
four-fermion contact interactions at
small energies compared to the compositeness
 scale $\Lambda$ via constituent 
exchanges, and this induces a correction of order $s/\Lambda^
{2}$ to the
QCD prediction of jet production \cite{r2}. The correction is 
negligible 
 at small energies, but 
becomes significant at high $p_{T}$.
This behavior agrees qualitatively with the observed inclusive 
jet
cross section.  We assume here that only quarks are composite 
and
gauge fields are elementary.

The CDF fit of the data using the tree level amplitudes from 
the 
effective Lagrangian by Eichten, Hinchliffe, Lane and Quigg 
(EHLQ) \cite{r3}
 with  $SU(2)_{L}$ doublet  quarks gives the compositeness scale 
 $\Lambda \approx 1.6 $ TeV. To go beyond the tree level analysis 
of 
 the data, we need  the QCD one-loop corrected amplitudes for 
$q_{i}q_
{j}
 \rightarrow q_{i}q_{j}$.
The leading QCD correction to the amplitudes arising from the 
QCD 
interaction has been known \cite{r4,r5,r6}, and so only the 
one-loop correction 
to 
the terms arising from the four-fermion contact interactions 
needs to 
be computed.

In this paper, we calculate the one-loop QCD correction to $q_
{i}q_{j} 
\rightarrow q_{i}q_{j} $ in the EHLQ effective Lagrangian, using
the framework of 
Kunszt,et.al. for one-jet inclusive cross section \cite{r6,r7}, 
and also
discuss QCD corrections in Drell-Yan process.
Section 2 through 6 is devoted to QCD corrections for the inclusive 
one-jet cross section. 
 In sec. 2 and  3, we review the EHLQ Lagrangian, and give the 
squared amplitudes at tree-level, and in sec.4 discuss the
ultraviolet divergence and summarize
the short distant effects of loop corrections. In sec.5 we briefly 
review
the method by Kunszt,et.al. for one-loop inclusive jet cross 
section,
and  give our result in sec.6. Details of QCD calculation may 
be found
in the appendix.  Finally in sec. 7 we discuss
QCD corrections in Drell-Yan process.

\section{Effective action}

A typical term of the 
helicity conserving effective interactions of composite quarks 
at low 
energies compared to the compositeness scale can be written 
in the form 
of 
current product
\be
{\cal L}_{int}(0)=g_{0}^{2}\eta(\mu,\Lambda) \int J_{\mu}^{R}(
\mu,x) J_
{\nu}^{R}(\mu,0) 
D_{\mu \nu}(x,\Lambda) d^{4} x,
\label{e1}
\ee
where $\eta(\mu, \Lambda)$ and $J_{\mu}^{R}$ are
 renormalized effective 
coupling and
generic quark current respectively.
The constituent exchange between currents is represented by 
$D_{\mu \nu}(x,\Lambda)$ which 
is assumed to satisfy
\bear
D_{\mu \nu}(x,\Lambda) &=& \Lambda^{2} g_{\mu \nu} D( x \Lambda), 
\\
D_{\mu \nu}(x,\Lambda) &\rightarrow & g_{\mu\nu}
 \frac{1}{\Lambda^{2}} \delta^{\left(4\right)}(x)
\hspace{.2in} \mbox{for}\hspace{.2in} x \Lambda  \gg 1.
\label{e2}
\eear
The $\Lambda$ in $D_{\mu \nu}$ is a cutoff that determines the 
interaction range of 
the constituent exchanges. The relation between $\eta(\mu,\Lambda),
J_{\mu}^{R}$ and the corresponding bare quantities  depends 
on
the fundamental dynamics at the compositeness scale. However,
this model  dependence does not cause any problem in calculating
QCD corrections at low energies
 because any ambiguity arising from the lack of knowledge
on how the currents and couplings are
renormalized can be absorbed in the coupling $\eta$ which is
supposed to be determined experimentally.
With (\ref{e2}),  (\ref{e1}) becomes at tree level
\be
\frac{g_{0}^{2} \eta}{\Lambda^{2}}\, J_{\mu}\,J^{\mu}.
 \ee

The most general helicity conserving, $SU(3)_{c} \times SU(2)_
{L} \times
U_{Y}(1)$ symmetric low-energy effective Lagrangian of up and 
down quarks 
by EHLQ is
\bear
{\cal L}_{EHLQ} &=& g^{2} \left(\frac{g_{0}^{2}}{2 g^{2} \Lambda^
{2}}\right) 
\left\{ \eta_{0} \bar{q}_{L} \gamma^{\mu} 
\bar{q}_{L} \bar{q}_{L} \gamma^{\mu} q_{L} +\eta_{1} \bar{q}_
{L} \gamma^
{\mu}\frac{\tau^{a}}{2} q_{L} \bar{q}_{L} \gamma^{\mu}\frac{\tau^
{a}}{2} 
q_{L} \right. \nonumber \\
                & & +\eta_{u} \bar{q}_{L} \gamma^{\mu} q_{L} 
\bar{u}_{R} 
\gamma^{\mu} u_{R} + 
                    \eta_{d} \bar{q}_{L} \gamma^{\mu} q_{L} 
\bar{d}_{R} 
\gamma^{\mu} d_{R} \nonumber \\
                & & + \eta_{8u} \bar{q}_{L} \gamma^{\mu} \frac{\lambda^
{a}}{2} q_{L} \bar{u}_{R} \gamma^{\mu} \frac{\lambda^{a}}{2} 
u_{R} + 
                   \eta_{8d} \bar{q}_{L} \gamma^{\mu} \frac{\lambda^
{a}}{2} 
q_{L} \bar{d}_{R} \gamma^{\mu} \frac{\lambda^{a}}{2} d_{R}  
                   \nonumber \\
                & & +\eta_{uu} \bar{u}_{R} \gamma^{\mu} u_{R} 
\bar{u}_
{R} \gamma^{\mu} u_{R} + \eta_{dd} 
                \bar{d}_{R} \gamma^{\mu} d_{R} \bar{d}_{R} \gamma^
{\mu} 
d_{R} \nonumber \\
                & & \left.+ \eta_{ud} \bar{u}_{R} \gamma^{\mu} 
u_{R} \bar{d}_
{R} \gamma^{\mu} d_{R} + 
                    \eta_{ud}' \bar{u}_{R} \gamma^{\mu} d_{R} 
\bar{d}_
{R} \gamma^{\mu} u_{R} \right\},
\label{e4} \eear
where $q_{L}=(u_{L},d_{L}).$ 
We inserted in (\ref{e4}) the strong coupling $g^{2}$ explicitly
to make the tree amplitudes of QCD and contact terms  be
formally in the same order in the QCD coupling. For convenience,
in the rest of the paper we absorb the factor
\be
\frac{g_{0}^{2}}{2 g^{2} \Lambda^{2}}
 \ee  
into the coupling $\eta$'s.
We also assume here all quarks are massless . Then because of 
the 
$SU_{L}(2)$ symmetry, there are only seven independent helicity
amplitudes for $q_{i}q_{j} \rightarrow q_{i}q_{j}$. They are: 
$u_{L} d_
{L} \rightarrow u_{L} d_{L}, u_{L} u_{L} \rightarrow u_{L} u_
{L},
u_{R} d_{L} \rightarrow u_{R} d_{L}, u_{L} d_{R} \rightarrow 
u_{L} d_{R}, 
u_{R} u_{R} \rightarrow u_{R} u_{R}, 
d_{R} d_{R} \rightarrow d_{R} d_{R}, $ and $ u_{R} d_{R} \rightarrow 
u_
{R} d_{R}. $
The amplitude for $d_{L} d_{R} \rightarrow d_{L} d_{R} $, for 
example, 
is 
identical to that of 
$u_{L} d_{R} \rightarrow u_{L} d_{R}$ because of the $SU(2)_
{L}$ symmetry. 
In the following,
we calculate these seven amplitudes to one-loop order in QCD.

\section{Tree amplitudes}

The  tree level amplitudes for the helicity channels in  the 
EHLQ effective 
action are given in the appendix. 
Note that we follow the notation in ref. \cite{r7,r12} for the 
helicity 
amplitude and spinor algebra.

The  squared amplitudes--- color and spin averaged ---for quark 
channels 
are 
\bear
\left|{\cal A}(ud \rightarrow ud)\right|^{2}&=&
\left|{\cal A}(\bar{u}\bar{d} \rightarrow \bar{u}\bar{d})\right|^
{2}= 
\nonumber \\
& & g^{4} \left[ \frac{4}{9} \frac{s^{2} +u^{2}}{t^{2}} + u^
{2}
\left( \eta_{u}^{2} +\eta_{d}^{2} +\frac{2}{9} ( \eta_{8u}^{2}+\eta_
{8d}^
{2})\right)\right.
\nonumber \\
& &+ s^{2} ( 4 \eta_{0}^{2} +\frac{2}{3} \eta_{0} \eta_{1} +\frac{11}{12} 
\eta_{1}^{2}
+\frac{2}{3} \eta_{ud} \eta_{ud}' + \eta_{ud}^{2} + \eta_{ud}'^
{2}) \nonumber 
\\
& &+ \frac{8}{9} \frac{s^{2}}{t} ( \eta_{1} + \eta_{ud}') + 
\left.
\frac{4}{9} \frac{u^{2}}{t} ( \eta_{8u} + \eta_{8d})\right],
 \eear
\bear
\left|{\cal A}(uu \rightarrow uu)\right|^{2}&=&
\left|{\cal A}(\bar{u}\bar{u} \rightarrow 
\bar{u}\bar{u})\right|^{2}= \nonumber \\
 & &g^{4} \left[ \frac{4}{9}\left( \frac{s^{2} +u^{2}}{t^{2}} 
+
\frac{s^{2} +t^{2}}{u^{2}} -\frac{2 s^{2}}{3 t u} \right)\right.
\nonumber \\
&&+ \frac{4}{9}\left( \frac{s^{2}}{t}+\frac{s^{2}}{u} \right)
( 4 \eta_{0} + \eta_{1}+ 4 \eta_{uu}) +  \frac{8}{9}
 \eta_{8u} \left( \frac{u^{2}}{t}+\frac{t^{2}}{u} \right) \nonumber 
\\
&&+\left. \frac{2 s^{2}}{3} ( 16 \eta_{0}^{2} + 8
 \eta_{0} \eta_{1} + \eta_{1}^{2} +16 \eta_{uu}^{2}) +
 2 (u^{2}+t^{2})( \eta_{u}^{2} + \frac{2}{9} \eta_{8u}^{2})
\right],
\label{e66} \eear
and using  the crossing symmetry
\be
\left|{\cal A}(u\bar{d} \rightarrow u\bar{d})\right|^{2}= 
\left|{\cal A}(\bar{u}d \rightarrow \bar{u}d)\right|^{2}=
\left|{\cal A}(ud \rightarrow ud)\right|^{2}( s \leftrightarrow 
u),
\ee
\be
\left|{\cal A}(\bar{u}\bar{u} \rightarrow \bar{d}\bar{d})\right|^
{2}= 
\left|{\cal A}(\bar{d}\bar{d} \rightarrow \bar{u}\bar{u})\right|^
{2}=
 \left|{\cal A}(ud \rightarrow ud)\right|^{2}
( s \rightarrow u, t \rightarrow s, u \rightarrow t),
\ee
\be
\left|{\cal A}(u\bar{u} \rightarrow u\bar{u})\right|^{2}=
\left|{\cal A}(uu \rightarrow uu)\right|^{2}( s \leftrightarrow 
u),
\label{e100}\ee
with
\be
s=(p_{1}+p_{2})^{2},\, t=(p_{1}-p_{4})^{2},\, u=(p_{1}-p_{3})^
{2}.
\ee
For the channels involving d-quarks only, we can obtain
 the squared amplitudes 
by replacing $\eta_{uu}, \eta_{8u}$  with
$\eta_{dd},\eta_{8d}$ respectively in (\ref{e66}),(\ref{e100}).

Putting back the factor
\be
\frac{g_{0}^{2}/{4 \pi}}{2 \alpha_{s} \Lambda^{2}}
 \ee
into each $\eta$'s in the above equations and keeping only $
\eta_{0}$
we can recover the formulas in \cite{r3} with
\be
g_{0}^{2}/{ 4 \pi}=1.
 \ee
Note that we have corrections for typos 
in (8.13), (8.15) in \cite{r3}.

\section{Ultraviolet divergence}

The one-loop Feynman diagrams for $q_{i}q_{j} \rightarrow q_
{i}q_{j}$ 
in the EHLQ effective
Lagrangian  are given in Fig.1.  In the massless limit, the 
one-loop
self energy diagrams for fermions in the dimensional regularization 
vanish \cite{pecc}, and (g),(h) in Fig.1  are the  UV
counter terms for current renormalization arising from the fermion 
self-energy
diagrams and  (a),(b).
(g),(h) are required only for color 
 octet currents, and the counter terms
 for the conserved, color singlet
currents vanish. We do not include in our calculation
 the one-loop diagram in Fig.2(b) because
its finite part coming from the small momentum region ($\sim 
\sqrt{s}$)
is suppressed by a factor  $s/\Lambda^{2}$ relative to those 
in Fig.1. 
This
can be easily seen from the fact that the four-fermion-gluon 
vertex
in the diagram is given by the effective Lagrangian
\be
 \sim \frac{1}{\Lambda^{4}}  f_{abc}\, J_{\mu}^{a}\, J_{\nu}^
{b} F_{\mu 
\nu }^{c}
 \ee
represented in Fig.2(a). Here $J_{\mu}^{a} $  denote color octet
currents. The contribution from the large momentum region
$(\gg \sqrt{s})$, which is dependent on the fundamental dynamics 
at the
compositeness scale,
is independent of external momenta, and renormalizes only the 
couplings
of the contact terms. Since we are not interested in how the 
$\eta$'s 
are renormalized, we can completely exclude this diagram. 

There are also penguin diagrams (Fig.3). Although they are not 
one-loop
QCD corrections, it is easy to see that they induce form factors
in the quark-gluon vertices  that are formally of same order 
of magnitude 
as the
QCD one-loop corrections. The form factors induced by the penguin 
diagrams
assume the form
\bear
F(q^{2}) &=& 1+ \frac{\eta g_{0}^{2}}{(4 \pi)^{2}} 
C( \ln(-q^{2}) )\left( \frac{q^{2}}{\Lambda^{2}} \right) \nonumber 
\\
          &\approx& 1 + \bar{C} \frac{\eta g_{0}^{2}}{(4 \pi)^
{2}} \left(
 \frac{q^{2}}{\Lambda^{2}} \right),
\label{e300}
\eear
where $ C\left( \ln(-q^{2})\right) $ is a  model dependent function 
of $O(1)$.  In the last step we replaced  the function  $ C 
$ with its average value $\bar{C}$ in the 
momentum range of interest. Thus penguin diagrams introduce 
new free parameters in the amplitudes. This form factor effect from
penguin diagrams  may be combined  with that  in the vector boson propagators \cite{drell, r3}. 

Now in general, (a)-(f) have ultraviolet divergences as well 
as soft and
collinear divergences. For one-jet inclusive cross section, 
the soft and collinear divergences are cancelled by those from 
$2 \rightarrow 
3$
process, which will be reviewed in more detail in the next section.
The UV divergences in (a), (b), as mentioned before, are cancelled
by the counter terms (g),(h). We assume the counter terms are 
given in 
the
$\overline{MS}$ scheme. The scheme dependence of the counter 
terms is 
absorbed
in the coupling $\eta$'s to render the physical amplitudes scheme
independent. The UV divergences in (c)-(f) arise from the approximation
in (\ref{e2}). If we insert $D_{\mu \nu}(x,\Lambda)$ between 
the currents, 
the 
diagrams would be finite with logarithmic terms of order $\alpha_
{s} \log(
\Lambda)$
from the short distance region. The scale $\Lambda$ plays the 
role of
an UV cutoff. The logarithmic terms can be summed to all orders 
in QCD
in the leading log approximation by applying RG-improved operator
product expansion to (\ref{e1}) \cite{r8,r9,r10}. Applying OPE 
to $J_{\mu}^
{R}(x) J_{\nu}^{R}(0)$,
\be
\int J_{\mu}^{R}(x) J_{\nu}^{R}(0) D_{\mu \nu}(x,\Lambda) dx 
=
\sum_{i} c_{i}(\mu/\Lambda,\alpha(\mu)) O_{i}^{R}(0),
\label{e7}
 \ee
with $c_{i}$ satisfying
\be
\left( \mu^{2} \frac{\partial}{\mu^{2}} + \beta(\alpha) \frac{\partial}{
\partial \alpha} + \tilde{\gamma_{i}} (\alpha)\right) c_{i}(
\mu/\Lambda,\alpha(
\mu))=0,
 \ee
where 
\be
  \tilde{\gamma_{i}} =2 \gamma_{J}-\gamma_{i},
   \ee
and $   \gamma_{J},\gamma_{i}  $ are the anomalous dimensions
of the current and the operator $O_{i}$ respectively, and 
\bear
\beta(\alpha)&=&\mu^{2} \frac{\partial}{\partial \mu^{2}} \alpha=
-\beta_{0} \alpha^{2} ( 1 + O(\alpha)), \nonumber \\
\beta_{0}&=& \frac{1}{4 \pi} ( 11 - \frac{2}{3} N_{f}),
 \eear
\be
 \tilde{\gamma_{i}}(\alpha)= \tilde{\gamma_{i}}^
{\left(1\right)} \alpha + O( \alpha^{2}).
 \ee
Note
that $\tilde{\gamma_{i}}$ arises only from the UV divergences 
in diagram
(c)-(f).

Integrating the RG equation, 
\bear
  c_{i}(\mu/\Lambda,\alpha(\mu)) &=& c_{i}(\alpha(\Lambda)) 
\exp\left(
\int
  _{\alpha(\mu)}^{\alpha(\Lambda)} \frac{\tilde{\gamma}_{i}(
\alpha)}{\beta(
\alpha)}
  d \alpha\right), \nonumber \\
                             &\approx& c_{i}^{\left(0\right)} 
L_{i}(\alpha(
\mu),\alpha(\Lambda))
\label{e12}
 \eear
where
\be 
L_{i}(\alpha(\mu),\alpha(\Lambda)) = \left( \frac{\alpha(\mu)}{\alpha(
\Lambda)}
\right)^{\frac{ \tilde{\gamma_{i}}^{\left(1\right)}}{\beta_{0}} 
}.
 \ee 
The constants $c_{i}^{\left(0\right)}$ can be determined from 
the tree-level 
amplitudes. Substituting (\ref{e12}) into (\ref{e7}), the one-loop 
effective
Lagrangian for (\ref{e1})  should read
\be
{\cal L}_{int}(0) = \sum g_{0}^{2} \eta (\mu, \Lambda) c_{i}^
{0}
L_{i} O_{i}^{R}.
\label{e14} \ee
In computing the matrix element $<\!f| O_{i}^{R}|i\!>$, we take 
the
$\overline{MS}$ subtraction scheme for the UV divergences. The 
scheme
dependence in the matrix element is compensated by that of $c_
{i}$
to make physical amplitudes scheme independent.

Since the EHLQ effective action at one-loop must assume the 
same form
as in the tree-level action, the short distance effect in diagram 
(c)-(f)
results in as a mixing among $\eta$'s with an appropriate scaling 
by $L_
{i}$.
Let us first consider $u_{L}d_{L} \rightarrow u_{L}d_{L} $ process.
The effective Lagrangian for this process is
\bear
{\cal L}_{u_{L}d_{L}} &=& ( 2 \eta_{0} - \frac{\eta_{1}}{2}) 
O_{1} + \eta_
{1} O_{2}  \nonumber \\
&=&( 2 \eta_{0} + \frac{\eta_{1}}{2}) O_{+} + 
( \frac{3 \eta_{1}}{2} -2 \eta_{0}) O_{-},
 \eear
where
\bear
O_{1} &=& \bar{u}_{L} \gamma^{\mu} u_{L} \bar{d}_{L} \gamma^
{\mu} d_{L}, 
  \nonumber \\
O_{2} &=& \bar{u}_{L} \gamma^{\mu} d_{L} \bar{d}_{L} \gamma^
{\mu} u_{L},
 \eear
and
\be
O_{\pm}=\frac{1}{2} (O_{2}\pm O_{1}).
 \ee

Because $O_{+} (O_{-})$ is symmetric (anti-symmetric) under 
interchange
between up and down quarks and separately up and down anti-quarks,
$O_{\pm}$ are multiplicatively renormalized to all orders in 
QCD, and
their anomalous dimensions at one-loop are given by \cite{r10}
\be
\gamma^{\left(1\right)}_{\pm}= \mp \frac{3}{4 \pi N_{c}} ( N_
{c}\mp 1).
 \ee
According to (\ref{e14}), the effective action at one-loop is 
\bear
{\cal L}_{u_{L}d_{L}} \rightarrow & & 
( 2 \eta_{0} + \frac{\eta_{1}}{2})L_{+} O_{+}^{R} + ( \frac{3 
\eta_{1}}{2} 
-2 \eta_{0}) 
L_{-} O_{-}^{R} \nonumber \\ 
                         &=& 
\frac{1}{2} \left( (2 \eta_{0} +\frac{\eta_{1}}{2}) L_{+} -
( \frac{3 \eta_{1}}{2} -2 \eta_{0})L_{-} \right) O_{1}^{R} \nonumber 
\\
                         &+& 
\frac{1}{2} \left( (2 \eta_{0} +\frac{\eta_{1}}{2}) L_{+} +
( \frac{3 \eta_{1}}{2} -2 \eta_{0})L_{-} \right) O_{2}^{R}, 
 \eear
with
\be
L_{\pm}(\alpha(\mu),\alpha(\Lambda)) = \left( \frac{\alpha(\mu)}{\alpha(
\Lambda)}
\right)^{-\frac{ \gamma_{\pm}^{\left(1\right)} } {\beta_{0}} 
}.
\label{e20}
\ee
Thus at one-loop order we have to replace $\eta_{0},\eta_{1}$ 
by
$\bar{\eta}_{0},\bar{\eta}_{1}$ defined by
\be
\left( \begin{array}{c}
         \bar{\eta}_{0} \\
         \bar{\eta}_{1}
         \end{array} \right) =\left( \begin{array}{cc}
                               \frac{3}{4} L_{+} +\frac{1}{4} 
L_{-} &
                               \frac{3}{16}(L_{+}-L_{-}) \\
                               L_{+}-L_{-} & \frac{1}{4}L_{+} 
+\frac{3}{4}
                               L_{-} 
                               \end{array} \right)
\left( \begin{array}{c}
         \eta_{0} \\
         \eta_{1}
         \end{array} \right).
\label{5171}
 \ee

Let us now consider $u_{L} d_{R} \rightarrow u_{L} d_{R}$ which 
involves 
a color octet
current. The Lagrangian for this process is
\be
{\cal L}_{u_{L} d_{R}}= \eta_{d} O_{1} + \eta_{8d} O_{2}
 \ee
where $O_{i}$ now are defined by
\bear
O_{1} &=& \bar{u}_{L} \gamma^{\mu} u_{L} \bar{d}_{R} \gamma^
{\mu} d_{R} 
 \nonumber \\
O_{2} &=& \bar{u}_{L} \gamma^{\mu} \frac{\lambda^{a}}{2} 
u_{L} \bar{d}_{L} \gamma^{\mu} \frac{\lambda^{a}}{2} d_{R}.
 \eear
Unlike in the previous example, in this case there is no simple 
symmetry argument to find  multiplicatively renormalized 
operators, and so we are going to diagonalize the one-loop 
mixing matrix explicitly.

{}From the UV divergence part in (a)-(f), we have 
\bear
\left( \begin{array}{c}
         O_{1} \\
         O_{2}
         \end{array} \right)^{B} = Z 
\left( \begin{array}{c}
         O_{1} \\
         O_{2}
         \end{array} \right)^{R},
 \eear
where
\bear
Z= \left(\begin{array}{ll}
1 & 
\frac{6 g^{2}}{(4 \pi)^{2}} \frac{1}{\epsilon}    \\
\frac{3 (N_{c}^{2}-1) g^{2}}{2 N_{c}^{2} (4 \pi)^{2}} \frac{1}{\epsilon} 
 &
1+\frac{3 (N_{c}^{2}-2) g^{2}}{N_{c} (4 \pi)^{2}} \frac{1}{\epsilon} 
\end{array} \right)
 \eear
with $ \epsilon=\frac{1}{2} (4-n)$. The one-loop anomalous dimension 
of $O_{i}$ is then
\be
\Gamma  = 
\frac{\alpha_{s}}{
4 \pi} \left( \begin{array}{cc}
          0 & 6 \\
          \frac{4}{3} & 7 
          \end{array} \right).
 \ee
Diagonalizing $\Gamma$, we have
\be
L \Gamma L^{-1} = 
\frac{\alpha_{s}}{
4 \pi} \left( \begin{array}{cc}
         -1  & 0 \\
          0 & 8
          \end{array} \right)
 \ee
with
\be
L=\left( \begin{array}{cc}
          1 & -\frac{3}{4} \\
          \frac{4}{27} & \frac{8}{9} 
          \end{array} \right).
 \ee
The one-loop anomalous dimension of octet current arising from 
(a)
in Fig.1 is
\be
\gamma_{j_{8}}^{\left(1\right)} = -\frac{N_{c}}{8 \pi }. 
 \ee
Thus at one-loop level,
\be
{\cal L}_{u_{L} d_{R}} \rightarrow \bar{\eta}_{d} O_{1}^{R} 
+ \bar{\eta}_
{8d} O_{2}^{R}
 \ee
where
\be
\left( \begin{array}{c}
         \bar{\eta}_{d} \\
         \bar{\eta}_{8d}
         \end{array} \right) =\left( \begin{array}{cc}
                                        c_{1} & d_{1} \\
                                        c_{2} & d_{2} 
                                        \end{array} \right)
\left( \begin{array}{c}
         \eta_{d} \\
         \eta_{8d}
         \end{array} \right),
\label{e30} \ee
\be
\left( \begin{array}{c}
         c_{1} \\
         c_{2}
         \end{array} \right) = L^{t} \left( \begin{array}{cc}
         L_{8+} & 0 \\
         0 & L_{8-} \end{array}
         \right) \left(L^{-1}\right)^{t} \left(\begin{array}{c}
         1 \\
         0 \end{array}
         \right),
 \ee
\be
\left( \begin{array}{c}
         d_{1} \\
         d_{2}
         \end{array} \right) = L^{t} \left( \begin{array}{cc}
         \tilde{L_{8}}_{+} & 0 \\
         0 & \tilde{L_{8}}_{-} \end{array}
         \right) \left(L^{-1}\right)^{t} \left(\begin{array}{c}
         0 \\
         1 \end{array}
         \right),
 \ee
and
\bear 
L_{8\pm} &=& \left(\frac{ \alpha(\mu)}{\alpha(\Lambda)}
\right)^{-\gamma_{8\pm}^{\left(1\right)}/\beta_{0}}, \nonumber 
\\
\tilde{L}_{8\pm} &=& \left(\frac{ \alpha(\mu)}{\alpha(\Lambda)}
\right)^{-(\gamma_{8\pm}^{\left(1\right)}-2 \gamma_{j_{8}}^{\left(
1\right)}
)/\beta_{0}}, \nonumber \\
 \eear
with
\bear
\gamma_{8+}^{\left(1\right)}&=& \frac{-1}{4 \pi}  \nonumber 
\\
\gamma_{8-}^{\left(1\right)}&=& \frac{8}{4 \pi}. 
 \eear

Similar calculation gives
\be
\left( \begin{array}{c}
         \bar{\eta}_{uu} \\
         \bar{\eta}_{dd}
         \end{array} \right) =\left( \begin{array}{cc}
                                L_{+}  &
                               0 \\
                               0 & 
                               L_{-} 
                               \end{array} \right)
\left( \begin{array}{c}
         \eta_{uu} \\
         \eta_{dd}
         \end{array} \right),
\label{5172}
 \ee
and
\be
\left( \begin{array}{c}
         \bar{\eta}_{ud} \\
         \bar{\eta}_{ud}'
         \end{array} \right) =\frac{1}{2}
         \left( \begin{array}{cc}
                               L_{+} + L_{-} &
                               L_{+}-L_{-} \\
                               L_{+}-L_{-} & L_{+} +
                               L_{-} 
                               \end{array} \right)
\left( \begin{array}{c}
         \eta_{ud} \\
         \eta_{ud}'
         \end{array} \right),
\label{5173}
 \ee
where $L_{\pm}$ are defined in (\ref{e20}).
The transformation for $\eta_{u}, \eta_{8u}$ can be obtained 
by
replacing $\eta_{d}, \eta_{8d}$ in (\ref{e30}) with
  $\eta_{u}, \eta_{8u}$
respectively.    The modified coupling $\bar{\eta}$'s should 
also 
be used in calculating the $2 \rightarrow 3$  tree-level amplitudes.
For notational convenience, we keep using $\eta$'s instead of 
the
modified couplings; however, in the rest of the paper, all $\eta$'s
should be understood as $\bar{\eta}$'s defined in (\ref{5171}), 
(\ref{e30}), (\ref{5172}), and (\ref{5173}).

\section{ Calculation framework}

When computing the one-jet inclusive cross section, we need 
the one-loop
QCD amplitudes for $2 \rightarrow 2$, not the squared 
amplitudes, since 
QCD and the composite model interaction act coherently. To use
the one-loop QCD helicity amplitudes for $2 \rightarrow 2$ by 
Kunszt,et.al. 
\cite{r6}
calculated in the 't Hooft Veltman scheme \cite{r11}, we also 
calculate 
the 
diagrams in Fig.1 in the 't Hooft Veltman scheme. Kunszt,et.al. 
also isolated the soft and collinear divergences in $2 \rightarrow 
2$ 
and
$2 \rightarrow 3$ processes, exposed explicitly the cancellation 
of
these divergences among them, and gave a complete prescription 
for
the one-jet inclusive cross section \cite{r6,r7}. In this section 
we briefly
review  the calculation scheme of  Kunszt,et.al., and identify 
terms to 
be 
calculated for the one-jet cross section. We follow the notation 
in
\cite{r6, r7} and readers should consult the references for 
more detailed
discussions.

The one-jet inclusive cross section 
\be
I = \frac{ d \sigma_{jet}}{dp_{J} dy_{J}}
 \ee
can be written as
\be
I= I(2\rightarrow 2) + I(2\rightarrow 3),
 \ee
where
\bear
I(2\rightarrow 2) &=& \frac{1}{2!} \int d \rho_{2} \frac{d \sigma(
2 \rightarrow 
2)}{d
\rho_{2}} S_{2}(p_{1}^{\mu},p_{2}^{\mu}) \nonumber \\
I(2\rightarrow 3) &=& \frac{1}{3!} \int d \rho_{3} \frac{d \sigma(
2 \rightarrow 
3)}{d
\rho_{3}} S_{3}(p_{1}^{\mu},p_{2}^{\mu},p_{3}^{\mu})
 \label{5177}
\eear
and
\bear
d \rho_{2} &=& dy_{1} d p_{2} d y_{2} d \phi_{2}  \nonumber 
\\
d \rho_{3} &=& dy_{1} d p_{2} d y_{2} d \phi_{2} 
d p_{3} d y_{3} d \phi_{3}.
 \eear
$S_{2},S_{3}$ define a jet algorithm, and $p_{i},y_{i}$ denote
transverse momenta and pseudo-rapidities of the partons respectively.

$I(2\rightarrow 2)$ can be divided into singular $( \sim 1/\epsilon^
{p} 
) $
and nonsingular parts
\be
I( 2\rightarrow 2) = I( 2\rightarrow 2)_{S} +I( 2\rightarrow 
2)_{NS}
\label{e41} \ee
with 
\be
I( 2\rightarrow 2)_{NS}= \frac{\alpha_{s}^{2}}{2 s^{2}} \int 
d \rho_{2} 
p_{2} \sum_{\bf a}
L_{AB} \left( \psi^{\left(4\right)}({\bf a,p}) +\frac{\alpha_
{s}}{2 \pi} 
\psi^{\left(6\right)}_{NS}({\bf a,p})\right) S_{2}(p_{1}^{\mu},p_
{2}^{\mu}),
 \ee
where
\be
L_{AB}= \frac{f_{A}(a_{A},x_{A})  f_{B}(a_{B},x_{B}) }{w(a_{A}) 
x_{A}
w(a_{B}) x_{B} },
 \ee
and $ {\bf a} =(a_{A},a_{B},a_{1},a_{2})$ for parton flavors, 
and
${\bf p} =(p_{A},p_{B},
p_{1},p_{2})$.
Indices $A,B$ and $1,2$  
 denote the initial state and the final
state partons respectively.   $\psi^{\left(4\right)}$ is
the Born amplitude squared and $\psi^{\left(6\right)}_{NS}$
is the nonsingular part of $\psi^{\left(6\right)}$ defined in
\be
\sum_{\textstyle {
  colors \atop
  spins }} \left| {\cal A}(a_{A}+a_{B} \rightarrow a_{1}+a_{2})
  \right|^{2}= g^{4} \left( \psi^{\left(4\right)}({\bf a,p}) 
+
  2 g^{2} c_{\Gamma} \left( \frac{\mu^{2}}{Q_{ES}^{2}}\right)^
{\epsilon}
  \psi^{\left(6\right)}({\bf a,p}) + O(g^{4})\right) 
  \label{e44} \ee
where $Q_{ES}$ is an arbitrary scale introduced by Ellis and 
Sexton \cite{
r4}, and
\be
c_{\Gamma}=\frac{1}{(4 \pi)^{2-\epsilon}} \frac{
\Gamma(1-\epsilon)^{2} \Gamma(1+\epsilon)}{\Gamma(1-2 \epsilon)}.
 \ee
The singular part $I(2\rightarrow 2)_{S}$ depends only on
 the Altarelli-Parisi
functions and the tree-level amplitudes in four dimensions, 
$ \psi^{\left(4\right)}({\bf a,p})$ and $ \psi_{mn}^{\left(
4,c\right)}(
{\bf a,p} )$, 
with the latter defined by
\be
 \psi_{mn}^{\left(4,c\right)}({\bf a,p}) = \frac{-2}{g^{4}} 
 T^{a}_{c_{\bar{m}} c_{m}} T^{a}_{c_{\bar{n}} c_{n}} \prod_
{i\neq m,n}
 \delta_{c_{\bar{i}}c_{i}} 
 {\cal A}^{\left(0\right)}_{c_{A}c_{B} c_{1}c_{2}}(2 \rightarrow 
2)
 {\cal A}^{\left(0\right)*}_{c_{\bar{A}}c_{\bar{B}} c_{\bar{1}}c_
{\bar{2}
 }}(2 \rightarrow 2).
 \label{e46} \ee
For $ {\bf a} =( q_{i},q_{j},q_{i},q_{j}) $, 
 $T^{a}= \lambda^{a}/2$ for the final state quarks and
$T^{a}= -\left(\lambda^{a}\right)^{t}/2$ for the initial state 
quarks.

Similarly $ I( 2\rightarrow 3)$ can be divided into singular 
and 
nonsingular parts
\bear
I( 2\rightarrow 3) &=& \left[ I( 2\rightarrow 3)- \sum_{n} I_
{n}'(2 \rightarrow 
3)\right]
+  \sum_{n} I_{n}'(2 \rightarrow 3)  \nonumber \\
&=& I_{finite}(2\rightarrow 3) + \sum_{n} I_{n}'(2 \rightarrow 
3)
\label{e47} \eear
where $n$ runs over $A,B,1,2$. The soft and collinear 
divergences are
contained in $ I_{n}'(2 \rightarrow 3) $, and $I_{n}'(2
 \rightarrow 3)$ is divergent
only when $p_{3}$ becomes soft or collinear
 to the parton $n$.
$I_{finite}(2 \rightarrow 3)$ is by construction
 well defined over all parton
phase space, depends only on the tree-level amplitudes
 in four dimensions, and so the phase-space integration
 can be done numerically.

Separating collinear divergence from soft divergence,
 $I_{n}'(2 \rightarrow 3)$ 
can be written as 
\be
I_{n}'(2 \rightarrow 3)= I_{n}^{soft}(2 \rightarrow 3) +
 I_{n}^{coll}(2 \rightarrow 3),
 \ee
with
\bear
I_{n}^{soft}(2 \rightarrow 3) &=&
 I_{n}^{soft}(2 \rightarrow 3)_{S} +
I_{n}^{soft}(2 \rightarrow 3)_{NS} \nonumber \\
I_{n}^{coll}(2 \rightarrow 3) &=&
 I_{n}^{coll}(2 \rightarrow 3)_{S} +
I_{n}^{coll}(2 \rightarrow 3)_{NS}, 
\label{e49} \eear
where the explicit form of  each term is given 
 in \cite{r6,r7}. The singular and nonsingular
 terms in (\ref{e49}) involve  
Altarelli-Parisi functions and only tree-level amplitudes in 
four dimensions. For example,
\be
I^{soft}_{2}(2 \rightarrow 3)_{NS} =
 \frac{\alpha_{s}^{3}}{4 \pi  s^{2}} \int d \rho_{2} p_{2}
\sum_{{\bf a}} L_{AB} \left[\psi_{2}^{soft} ({\bf a,p})\right]_
{NS}
 S_{2}(p_{1}^{\mu},p_{2}^{\mu})
 \ee
with 
\be
\left[{\psi_{2}^{soft}}\right]_{NS} = \sum_{m= A,B,1}
 \psi_{2m}^{\left(4,c\right)} \,\,
\tilde{T}_{2m},
 \ee
where 
 $ \tilde{T}_{2m}$ is a universal function of $
s_{ij}=(p_{i}+p_{j})^{2} $.

Adding (\ref{e47}) into (\ref{e41}), we have
\be
I = I(2 \rightarrow 2)_{NS} + \sum_{n}\left(I_{n}^
{soft}(2 \rightarrow 3)_{NS} +
I_{n}^{coll}(2 \rightarrow 3)_{NS}  \right) +
 I_{finite}(2\rightarrow 3)
\label{e52} \ee
with  complete cancellation of the singular terms.
{}From (\ref{e52}) we see that for the one-jet
 inclusive cross section,  we
need to calculate the tree-level amplitude
 $\psi^{\left(4\right)},
\psi_{mn}^{\left(4,c\right)}$ in four dimensions, and the
one-loop amplitudes $\psi_{NS}^{\left(6\right)}$
 of $\psi^{\left(6\right)}$
defined in (\ref{e44}).

\section {One-loop amplitudes}

The general form of the  amplitude 
for $q_{i}(p_{1}) q_{j}(p_{2})\rightarrow q_{k}(p_{3}) 
q_{l}(p_{4}) $ to one-loop for each helicity channel  can be
written as
\be
{\cal A} (p_{1}+p_{2} \rightarrow
 p_{3} + p_{4}) = g^{2} \tilde{c}\,
( A\, \delta_{li}\delta_{kj} +
 B\, \delta_{lj}\delta_{ki}  ),
\label{e105}
 \ee
where
\bear
A &=& A^{\left(0\right)} + 
g^{2} A^{\left(1\right)} \nonumber \\
B &=& B^{\left(0\right)} + g^{2} B^{\left(1\right)},
\label{e106}
 \eear
and
\be
A^{\left(i\right)} = A^{\left(i\right)}_{
QCD} +A^{\left(i\right)}_{cont}, \hspace{.2in}
B^{\left(i\right)} = B^{\left(i\right)}_
{QCD} +B^{\left(i\right)}_{cont}.
\label{e101}
\ee
$\tilde{A}^{\left(i\right)}_{QCD}, \tilde{B}^{\left(i\right)}_
{QCD}$
and $\tilde{A}^{\left(i\right)}_{cont}, \tilde{B}^{\left(i\right)}_
{con}$
are the tree and one-loop amplitudes arising from QCD and the 
fermion 
contact
interactions respectively, and  $ \tilde{c}$ is a channel dependent
spinor matrix element.

The amplitude squared for each helicity channel is then
\bear
\sum_{colors}|{\cal A}|^{2} &=& g^{4}
 |\tilde{c}|^{2} N_{c}^{2} \left\{
\left(A^{\left(0\right)}\right)^{2} +
 \left(B^{\left(0\right)}\right)^{2} +
 \frac{2}{N_{c}} A^{\left(0\right)} B^{\left(0\right)}
 + 2 g^{2} \left[ A^{\left(0\right)}
 Re(A^{\left(1\right)}) \right. \right. \nonumber \\
& &\left.\left.+B^{\left(0\right)}
 Re(B^{\left(1\right)}) + \frac{1}{N_{c}}\left(
A^{\left(0\right)} Re(B^{\left(1\right)})
 + B^{\left(0\right)}
 Re(A^{\left(1\right)}) \right)\right] \right\}.
\label{e62} \eear
Comparing (\ref{e62}) with (\ref{e44}), we have $ \psi^{\left(
4\right)},
\psi^{\left(6\right)}_{NS}$ in $q_{i}q_{j} \rightarrow q_{i}q_
{j}$ channel,
\bear
\psi^{\left(4\right)} &=& \sum_{spins} 
|\tilde{c}|^{2} N_{c}^{2} \left[
 \left(A^{\left(0\right)}\right)^{2} +
 \left(B^{\left(0\right)}\right)^{2} +
\frac{2}{N_{c}} A^{\left(0\right)} B^{
\left(0\right)} \right] \nonumber \\
\psi^{\left(6\right)}_{NS} &=& \sum_{spins}
 |\tilde{c}|^{2} N_{c}^{2} \left[ 
A^{\left(0\right)} Re(\tilde{A}^{\left(1\right)})\! 
+ \! B^{\left(0\right)} Re(\tilde{B}^{
\left(1\right)}) \!+\!\frac{1}{N_{c}} ( 
A^{\left(0\right)} Re(\tilde{B}^{\left(1\right)})
 + B^{\left(0\right)} Re(\tilde{A}^{
\left(1\right)}) ) \right],
 \eear
where
\bear
\tilde{A}^{\left(1\right)} &=& 
\left[ \left( \frac{Q_{ES}^{2}}{
\mu^{2}}\right)^{\epsilon}
\frac{1}{c_{\Gamma}} A^{\left(1\right)}
 \right]_{NS} \nonumber \\
\tilde{B}^{\left(1\right)} &=& \left[ 
\left( \frac{Q_{ES}^{2}}{\mu^{2}}\right)^{\epsilon}
\frac{1}{c_{\Gamma}} B^{\left(1\right)} \right]_{NS}.
\label{e107}
 \eear

For the one-loop amplitudes $A^{\left(1\right)}_{cont},
B^{\left(1\right)}_{cont}$, we
 calculate the diagrams in Fig.1 in the 't Hooft
 Veltman dimensional
regularization scheme in which the
 spins and momenta of internal
particles are defined in $n $ dimensions,
 while those of external particles
are defined in four dimensions.  
The calculation of the one-loop diagrams 
in the 't Hooft Veltman scheme is much simplified 
since we can treat
the Dirac $\gamma$ matrices as if they were defined in four
dimensions. To show this, let us consider, as an example, the 
diagram (a) in Fig.1 for $u_{L}(p_{1}) 
d_{L}(p_{2}) \rightarrow u_{L}(p_{4})
 d_{L}(p_{3})$. The diagram (a) is
proportional to
\bear
& & \int d^{n}k \frac{ <\!4\! - \!|\gamma^{\mu}( 
\not\! k +\not\! p_{4}) \gamma^{\alpha}
(1-\gamma_{5})(\not\! p_{1} + \not\! k ) 
\gamma^{\mu} |1\! - \! \!><\!3\! - \!
 |\gamma^{\alpha}|2\! - \! \!>}{
k^{2} ( k +  p_{4})^{2} ( k+p_{1})^{2}} \nonumber \\
&=&2\int_{0}^{1}d x\, x \int_{0}^{1} d y \int d^{n}k \frac{ 
<\!4\! - \!
 |\gamma^{\mu}(\not\! k +\not\! p
) \gamma^{\alpha}
(1-\gamma_{5})(  \not\! q + \not\! k )
 \gamma^{\mu} |1\! - \! \!><\!3\! - \! |
\gamma^{\alpha}|2\! - \! \!>}{
(k^{2} - x^{2} y (1-y) s_{14})^{3}}, 
\label{e53} \eear
where $p= -x y p_{1}+(1-x +x y)p_{4},$ and 
 $q= (1-x y) p_{1}-x(1-y)p_{4}.$
Writing $ \gamma^{\mu}$, defined 
in $n$-dimensions, as
\be
\gamma^{\mu} =\gamma_{\left(4\right)}^{\mu}
 + \underline{\gamma}^{\mu},
 \ee
where $\gamma_{\left(4\right)}^{\mu}$
 are the four dimensional
Dirac matrices and 
\bear
&&  \underline{\gamma}^{\mu} =0
  \hspace{.2in} \mbox{for}\hspace{.2in}
  \mu \leq 4, \nonumber \\
&&[ \gamma_{5},
\underline{\gamma}^{\mu}]=0, \hspace{.2in} \mbox{for}\hspace{.2in}
  \mu > 4, 
\eear
the numerator in the integrand in 
(\ref{e53}) can be written as
\bear
&& <\!4\! - \! |\gamma^{\mu}( 
\not k_{\left(4\right)} + \not\! p
) \gamma_{\left(4\right)}^{\alpha}(
  \not \! q + \not \! k_{\left(4\right)} ) \gamma^{\mu}_{\left
(4\right)}
 |1\! - \! \!><\!3\! - \! |\gamma_{
\left(4\right)}^{
\alpha}|2\! - \! \!> \nonumber \\
&& + <\!4\! - \! |\underline{\gamma}^{\mu}
 \not\underline{k}
\gamma_{\left(4\right)}^{\alpha}
\not\underline{k} \underline{
\gamma}^{\mu}|1\! - \! \!><\!3\! - \! | 
\gamma_{\left(4\right)}^{\alpha}|2\! - \! \!>
\label{e55} \eear
The second term in (\ref{e55})
 is $O(\epsilon^{4})$ and so it can be 
safely discarded because  the soft and 
collinear divergence is at most 
$O(1/\epsilon^{2})$. The four-dimensional Dirac algebra then 
gives the divergent term
\be
 <\!4\! - \! |\gamma_{\left(4\right)}^{\mu}
 \not k_{\left(4\right)}\gamma_{\left(4\right)}^{\alpha}
\not k_{\left(4\right)}
\gamma^{\mu}_{\left(4\right)}|1\! - \! \!>
<\!3\! - \! |\gamma_{\left(4
\right)}^{\alpha}|2\! - \! \!> 
 \ee
to
\be
\frac{8 k^{2}}{n} [12]<\!34\!>,
 \ee
where $k^{2}$ is defined in $n$-dimensions.
Then the integration over $k$ in $n$ dimensions can be
done in the standard way \cite{pecc}. 
 For other diagrams we can
similarly check that only the four
 dimensional $\gamma$-matrices contribute
to the loop diagrams.

As mentioned before, the counter
 terms  (g),(h) are nonvanishing
only for the color octet currents.
 For the octet currents, from the fermion 
self-energy diagrams and (a),(b), they are
given by
\be
\frac{N_{c}}{2} g^{2} \left(\frac{1}{
\epsilon}\right) c_{\Gamma}
\cdot {\cal A}_{tree},
 \ee
where $ {\cal A}_{tree}$  denotes the tree 
amplitude of the corresponding
contact term.
Also the UV divergences in (c)-(f) should be 
subtracted  in the $\overline{MS}$ 
scheme.
For each helicity channel, we give $
 \psi_{mn}^{\left(4,c\right)}, \tilde{c},
 A^{\left(0\right)}, \tilde{A}^{\left(1\right)},B^{
\left(0\right)}, \tilde{B}^{\left(1\right)}$ in 
appendix.

\section{Drell-Yan Process}

If composite quarks and leptons share common constituents,  
exchanges
between quarks and leptons of their common constituents  would 
give rise
to quark-lepton contact interactions at low energies. Then the 
signal 
from
these contact interactions may appear in Drell-Yan processes. 
In 
this
section we write down a general effective quark-lepton contact 
interactions
and consider their one-loop QCD corrections in Drell-Yan processes.

As in the EHLQ lagrangian, we consider a single family of quarks 
and leptons.
Including more fermion families should be straightforward. With 
the first
generation of fermions, the most general, helicity preserving, 
$SU(3)_
{QCD}
\times SU(2)_{L}\times U(1)_{Y}$ symmetric quark-lepton contact 
interactions
are
\bear
{\cal L}_{QL} &=& \frac{g_{0}^{2}}{2 \Lambda^{2}}
\left\{ \xi_{0} \bar{q}_{L} \gamma^{\mu} 
\bar{q}_{L} \bar{l}_{L} \gamma^{\mu} l_{L} +\xi_{1} \bar{q}_
{L} \gamma^
{\mu}\frac{\tau^{a}}{2} q_{L} \bar{l}_{L} \gamma^{\mu}\frac{\tau^
{a}}{2} 
l_{L} \right. \nonumber \\
                & & +\xi_{u} \bar{l}_{L} \gamma^{\mu} l_{L} 
\bar{u}_{R} 
\gamma^{\mu} u_{R} + 
                    \xi_{d} \bar{l}_{L} \gamma^{\mu} l_{L} \bar{d}_
{R} 
\gamma^{\mu} d_{R} + \xi_{e} \bar{q}_{L} \gamma^{\mu} q_{L} 
\bar{e}_{R} 
\gamma^{\mu} e_{R} \nonumber \\ 
                & & \left. +\xi_{ue} \bar{u}_{R} \gamma^{\mu} 
u_{R} \bar{e}_
{R} \gamma^{\mu} e_{R} + \xi_{de} 
                \bar{d}_{R} \gamma^{\mu} d_{R} \bar{e}_{R} \gamma^
{\mu} 
e_{R} + \left( \xi_{s}  \bar{q}_{L}^{i} d_{R} \bar{e}_
{R} l_{L}^{i} +  h.c. \right) \right\},
\label{e400} \eear
where $ l_{L}=(\nu,e)_{L}, \,\, q_{L}=(u,d)_{L}$. 
The last term in (\ref{e400}) is due to  scalar (or pseudo scalar)
exchanges and all the other terms are due to vector (or axial 
vector)
exchanges. 

For the massless fermions, no amplitudes with the  same fermion 
helicities in the
scalar exchange term arise in the standard model, and so the 
contact term of the scalar exchanges provides the leading amplitude 
in that
helicity channel. Therefore  we may keep the  amplitudes in 
the scalar
channel at tree-level, and consider one-loop QCD corrections 
only in the
vector channels.

To calculate the one-loop QCD corrections in the quark sector 
in Drell-Yan
process, we must consider virtual corrections in $q \bar{q}' 
\rightarrow l \bar{l'}
$ and the real gluon emission in $ q \bar{q}' \rightarrow l 
\bar{l'} G$ along with
$ q G \rightarrow q'l \bar{l}'$. 
Since these processes occur only in s-channel in the lepton 
momenta, the amplitudes 
for a given helicity
channel factorize into a flavor-independent part and a
flavor-dependent propagator part that also  includes the couplings 
on the quark
and lepton vertices.  The contact interactions thus  modify 
only the propagator part, and so  the one-loop QCD corrections 
in this model
are  essentially identical to those in  the standard model.
This allows us to write the cross section at 
parton level in terms of the corresponding cross section with 
a virtual photon
exchange  in the standard model 
\be
d \sigma^{\left(h\right)}( q \bar{q}' \rightarrow l \bar{l}') 
= d \sigma^{\left(h\right)
}_{\gamma^{*}}( 
q \bar{q} \rightarrow l \bar{l}) \cdot \left| \frac{D^{\left(
h\right)}(q \bar{q}' 
\rightarrow l \bar{l}') q^{2}}{Q_{q} Q_{l}}\right|^{2},
\label{401}
\ee
where $h, Q_{q,l}$ denote the helicities and charges of quarks 
and leptons
respectively  and
$ q^{2}$ is the invariant mass squared of the  leptons.
The helicity independence of $ d \sigma^{\left(h\right)}_{\gamma^
{*}}/ d Q^{2}$,
where $ q^{2}= Q^{2},$ of virtual photon exchange
allows us to write (\ref{401}) as
\be
\frac{d \sigma^{\left(h\right)}}{d Q^{2}}( q \bar{q}' \rightarrow 
l \bar{l}')
 = \frac{d \sigma_{ \gamma^{*}}}{d Q^{2}} (
 q \bar{q} \rightarrow l \bar{l}) \cdot \left| \frac{D^{\left(
h\right)}(q
 \bar{q}' \rightarrow l \bar{l}') q^{2}} {Q_{q} Q_{l}}\right|^
{2}.
\label{402}
\ee
Similarly for the $q G \rightarrow q'l\bar{l}'$,
\be
\frac{d \sigma^{\left(h\right)}}{d Q^{2}}( q G \rightarrow q' 
l \bar{l}')
 = \frac{d \sigma_{ \gamma^{*}}}{d Q^{2}} (
 q G \rightarrow q l \bar{l}) \cdot \left| \frac{D^{\left(h\right)}(
q
 \bar{q}' \rightarrow l \bar{l}') q^{2}} {Q_{q} Q_{l}}\right|^
{2}.
\label{403}
\ee 
From $d\sigma_{\gamma^{*}} / d Q^{2}$ in Altarelli, Ellis and 
Martinelli
 \cite{r100}, we finally have the Drell-Yan 
cross section for $ l \bar{l}'$ pair production, 
\bear
\frac{d \sigma^{DY}}{d Q^{2}} &=& \frac{1}{4}\frac{1}{36 \pi 
s Q^{2}} \int
_{0}^{1} \frac{d x_{1}}{x_{1}}\int_{0}^{1} \frac{d x_{2}}{x_
{2}} 
\sum_{f,f'}\left\{ \left[  q_{f}^{\left[1\right]}(x_{1}) 
\bar{q}_{f'}^{\left[2\right]}(x_{2}) + ( 1 \leftrightarrow 2)\right] 
\right.  
 \nonumber   \\
                  & &     \times  \left[ \delta(1-z) + 
\alpha_{s}(Q^{2}) \theta(1-z) (f_{q,DY}(z) -2 f_{q,2}(z))\right] 
\nonumber \\
                  & & \left.+ \left[( q_{f}^{\left[1\right]}(
x_{1}) + \bar{q}_{f'}^{\left[1\right]}(x_{1}))G^{\left[2\right]}(
x_{2}) +
 (1 \leftrightarrow 2) \right] 
                 \alpha_{s}(Q^{2}) \theta(1-z)(f_{G,DY}(z)
-f_{G,2}(z)) \right\} \nonumber \\ 
  & & \times \sum_{h}\left| D^{\left(h\right)}(q_{f} \bar{q}_
{f'}
 \rightarrow l \bar{l}') q^{2}\right|^{2},
\eear
with
\bear
f_{q,DY}-2 f_{q,2} &=& \frac{2}{3 \pi} \left[ \frac{3}{(1-z)_
{+}} -6-4 z+2 (1+z^{2})
\left(\frac{\ln(1-z)}{1-z}\right)_{+} + ( 1+\frac{4}{3} \pi^
{2} ) \delta(1-z)\right],
\nonumber \\
f_{G,DY}-f_{G,2} &=& \frac{1}{4\pi} \left[ \frac{3}{2} -5  z 
+\frac{9}{2}z^{2}
+ ( z^{2} +(1-z)^{2}) \ln (1-z) \right],
\eear
where $z= Q^{2}/x_{1}x_{2}s$ and $\sqrt{s}$ is the invariant 
mass of the incoming
hadron system.

From the standard model  lagrangian and  (\ref{e400}) we can 
read off the nonvanishing propagator part
\bear
&&D(u_{L}\bar{d}_{L} \rightarrow \nu_{L} \bar{e}_{L} ) = \frac{g^
{2}}{2 ( 
q^{2}-M_{w}^{2})} + \frac{g_{0}^{2}\xi_{1}}{4 \Lambda^{2}}, 
 \nonumber \\
&&D(u_{L}\bar{u}_{L} \rightarrow e_{L} \bar{e}_{L} ) = \left(
\frac{g}{
4 \cos\theta_{w}}\right)^{2} \frac{u_{-}l_{-}}{ q^{2}-M_{z}^
{2}} +
\frac{Q_{u}Q_{e}}{q^{2}} +\frac{g_{0}^{2}\xi_{0}}{2 \Lambda^
{2}} - \frac{g_{0}^{2}\xi_{1}}{8 \Lambda^{2}},  \nonumber \\ 
&&D(u_{L}\bar{u}_{L} \rightarrow e_{R} \bar{e}_{R} ) = \left(
\frac{g}{
4 \cos\theta_{w}}\right)^{2} \frac{u_{-}l_{+}}{ q^{2}-M_{z}^
{2}} +
\frac{Q_{u}Q_{e}}{q^{2}} +\frac{g_{0}^{2} \xi_{e}}{ 2 \Lambda^
{2}}, \nonumber \\
&&D(u_{R}\bar{u}_{R} \rightarrow e_{L} \bar{e}_{L} ) = \left(
\frac{g}{
4 \cos\theta_{w}}\right)^{2} \frac{u_{+}l_{-}}{ q^{2}-M_{z}^
{2}} +
\frac{Q_{u}Q_{e}}{q^{2}} +\frac{g_{0}^{2} \xi_{u}}{2 \Lambda^
{2}}, \nonumber \\ 
&&D(u_{R}\bar{u}_{R} \rightarrow e_{R} \bar{e}_{R} ) = \left(
\frac{g}{
4 \cos\theta_{w}}\right)^{2} \frac{u_{+}l_{+}}{ q^{2}-M_{z}^
{2}} +
\frac{Q_{u}Q_{e}}{q^{2}} +\frac{g_{0}^{2}\xi_{ue}}{2 \Lambda^
{2}}, \nonumber \\ 
&&D(d_{L}\bar{d}_{L} \rightarrow e_{L} \bar{e}_{L} ) = \left(
\frac{g}{
4 \cos\theta_{w}}\right)^{2} \frac{d_{-}l_{-}}{ q^{2}-M_{z}^
{2}} +
\frac{Q_{d}Q_{e}}{q^{2}} +\frac{g_{0}^{2}\xi_{0}}{2 \Lambda^
{2}} + \frac{g_{0}^{2}\xi_{1}}{8 \Lambda^{2}},  \nonumber \\
&&D(d_{L}\bar{d}_{L} \rightarrow e_{R} \bar{e}_{R} ) = \left(
\frac{g}{ 
4 \cos\theta_{w}}\right)^{2} \frac{d_{-}l_{+}}{ q^{2}-M_{z}^
{2}} +
\frac{Q_{d}Q_{e}}{q^{2}} +\frac{g_{0}^{2} \xi_{e}}{2 \Lambda^
{2}}, \nonumber \\
&&D(d_{R}\bar{d}_{R} \rightarrow e_{L} \bar{e}_{L} ) = \left(
\frac{g}{
4 \cos\theta_{w}}\right)^{2} \frac{d_{+}l_{-}}{ q^{2}-M_{z}^
{2}} +
\frac{Q_{d}Q_{e}}{q^{2}} +\frac{ g_{0}^{2}\xi_{d}}{2 \Lambda^
{2}}, \nonumber \\ 
&&D(d_{R}\bar{d}_{R} \rightarrow e_{R} \bar{e}_{R} ) = \left(
\frac{g}{
4 \cos\theta_{w}}\right)^{2} \frac{d_{+}l_{+}}{ q^{2}-M_{z}^
{2}} +
\frac{Q_{d}Q_{e}}{q^{2}} +\frac{g_{0}^{2}\xi_{de}}{2 \Lambda^
{2}}, 
\eear
where
$$
\begin{array}{lcrclcr}
u_{-}&=& 2 - \frac{8}{3} \sin^{2}\theta_{w};& \,\,\,\,\,\,& 
u_{+}&=& -\frac{8}{3} \sin^{2}\theta_{w},  \\
d_{-}&=& -2 + \frac{4}{3} \sin^{2}\theta_{w};& \,\,\,\,\,\, &
d_{+}&=& \frac{4}{3}  \sin^{2}\theta_{w},  \\
l_{-}&=& -2 + 4 \sin^{2}\theta_{w};& \,\,\,\,\,\, &
l_{+}&=& 4 \sin^{2}\theta_{w},
\end{array}
$$
and $g$ is the $SU(2)_{L}$ coupling constant.

Finally we would like to add a comment on calculating $ d \sigma^
{DY} / d p_{l} d y_{l}$ in the framework of Kunszt, et.al.,
where $p_{l}, y_{l}$ are the transverse momentum and rapidity 
of a designated lepton respectively. Since the essential difference 
of
Drell-Yan process from the one-jet inclusive production  is 
that the soft and
collinear divergences in Drell-Yan process  occur only in the 
initial state, 
all the formulae for the inclusive jet cross section can be
used  with  only minor modification.   First, since the final 
state leptons
in Drell-Yan process are identifiable,  the jet algorithm in 
(\ref{5177}) 
should be replaced  by
\be
S_{2}=2 ! \delta(p_{l}-p_{1}) \delta (y_{l}-y_{1}) , \hspace{.5in} 
S_{3}= 3 ! \delta(p_{l}-p_{1}) \delta (y_{l}-y_{1}), 
\label{5175}
\ee
and since soft and collinear divergences occur only in the initial 
state, 
the indices $m, n$ in (\ref{e46}), (\ref{e47}), and  (\ref{e52}) 
 should run only over the 
initial state. Also the restriction on $p_{3}$ that it is the 
smallest among
the final state parton momenta should be revoked.

 For the virtual corrections, we
need to consider only the diagram (a) in Fig.1, and its calculation 
goes
exactly as in the case of the one-jet inclusive cross section.
Details of the calculation and  
numerical analysis of the  Drell-Yan cross section along with 
the
one-jet inclusive cross section will be published elsewhere 
\cite{r101}.

\section{Acknowledgements}

I am grateful to E. Eichten for interesting me in this subject 
and
many valuable suggestions throughout the work. Also helpful
discussions with and comments by S. Parke and W. Giele are
gratefully acknowledged.

\newpage
\appendix
\section{Appendix} 

In this appendix we give the tree
 amplitudes $\psi_{mn}^{\left(4,c\right)},
 \tilde{c}, A_{cont}^{
\left(0\right)}, B_{cont}^{\left(0\right)}$,
 and the one-loop amplitudes $\tilde{A}_{cont}^{
\left(1\right)}\tilde{B}_
{cont}^{\left(1\right)} $ defined in (\ref{e46}),
(\ref{e105}),(\ref{e106}) and (\ref{e107}), 
and for completeness, 
corresponding terms in QCD for the following
  helicity channels: $ {\rm (i)}\,\,\, u_{L} d_{L}
 \rightarrow u_{L} d_{L},\,\, {\rm (ii)}\,\,\, u_{L}
 u_{L} \rightarrow u_{L} u_{L},
\,\,{\rm (iii)}\,\,\, u_{R} d_{L} \rightarrow u_{R}
 d_{L}, \,\,{\rm (iv)}\,\,\, u_{L} d_{R} \rightarrow 
u_{L} d_{R}, \,\,{\rm (v)}\,\,\, u_{R} u_{R}
 \rightarrow u_{R} u_{R}, 
\,\,{\rm (vi)}\,\,\, d_{R} d_{R} \rightarrow d_{R} d_{R},
 $ and $ \,\,{\rm (vii)}\,\,\, u_{R} d_{R}
 \rightarrow u_{R} d_{R}. $  All the momenta of external fermions
are assumed to have positive energies, and $s_{ij}=(p_{i}+p_
{j})^{2}.$

${\bf\tilde{c}:}$ 
$$
\begin{array}{ccc}
\hspace{.5in}&\tilde{c} &\hspace{1in} |\tilde{c}|^{2} \\ 
& & \\
\hspace{.2in}{\rm (i) }\hspace{.5in}& -2 i [12]<\!34\!>  &
\hspace{1in}  4 s_{12}^{2} \\
\hspace{.2in}{\rm (ii)}\hspace{.5in}& -2 i[12]<\!34\!>
   & \hspace{1in}  4 s_{12}^{2} \\
\hspace{.2in}{\rm (iii)}\hspace{.5in}& - 2 i [24]<\!13\!>
 & \hspace{1in}  4 s_{13}^{2} \\
\hspace{.2in}{\rm (iv) }\hspace{.5in}& -2 i [13]<\!24\!>
  & \hspace{1in}  4 s_{13}^{2} \\
\hspace{.2in}{\rm (v)  }\hspace{.5in}& -2 i [34]<\!12\!> 
 & \hspace{1in}  4 s_{12}^{2} \\
\hspace{.2in}{\rm (vi) }\hspace{.5in}&-2 i [34]<\!12\!>  
 & \hspace{1in}  4 s_{12}^{2} \\
\hspace{.2in}{\rm (vii)}\hspace{.5in}& -2 i [34]<\!12\!>
  & \hspace{1in}  4 s_{12}^{2} 
\end{array} \eqno (A.1)
$$

${\bf A^{\left(0\right)},B^{\left(0\right)}}$: For contact terms,
$$
\begin{array}{crr}
  \hspace{.5in} & A_{cont}^{\left(0\right)}&
 B_{cont}^{\left(0\right)} \\
 & & \\
\hspace{.2in}{\rm (i)}\hspace{.5in} & -2 \eta_{0} +\eta_{1}/2 
   & \hspace{.5in}      -\eta_{1}\\
\hspace{.2in}{\rm (ii)}\hspace{.5in} & -2( \eta_{0} +\eta_{1}/4)
   & \hspace{.5in}   -2( \eta_{0} +\eta_{1}/4)\\
\hspace{.2in}{\rm (iii)}\hspace{.5in} & -\eta_{u} +
 \eta_{8u}/{2 N_{c}}  & \hspace{.5in}  -\eta_{8u}/2\\
\hspace{.2in}{\rm (iv)}\hspace{.5in} &-\eta_{d} +
 \eta_{8d}/{2 N_{c}}    &\hspace{.5in}   -\eta_{8d}/2\\
\hspace{.2in}{\rm (v)}\hspace{.5in} &-2 \eta_{uu}     
       & \hspace{.5in}        -2 \eta_{uu}\\
\hspace{.2in}{\rm (vi)}\hspace{.5in} & -2 \eta_{dd}   
    & \hspace{.5in}       -2 \eta_{dd}\\
\hspace{.2in}{\rm (vii)}\hspace{.5in} & - \eta_{ud}  
  &   \hspace{.5in} - \eta_{ud}'  \\ 
   & & \\
\hspace{-.2in}\mbox{and for QCD,} 
&A_{QCD}^{\left(0\right)} & B_{QCD}^{\left(0\right)}\\
& & \\
\hspace{.2in}{\rm (i)}\hspace{.5in}& -1/{2 N_{c} s_{14}} &\hspace{.5in}
1/{2 s_{14}} \\
\hspace{.2in}{\rm (ii)}\hspace{.5in}& -1/{2 N_{c} s_{14}} +1/{2 
s_{13}} 
  &\hspace{.5in}  -1/{2 N_{c} s_{13}} +1/{2 s_{14}}  \\     
       \hspace{.2in}{\rm (iii)}\hspace{.5in} & -1/{2 N_{c} 
s_{14}}
&\hspace{.5in}      1/{2 s_{14}}\\
\hspace{.2in}{\rm (iv)}\hspace{.5in} & -1/{2 N_{c} 
s_{14}} &\hspace{.5in}     1/{2 s_{14}}\\
\hspace{.2in}{\rm (v)}\hspace{.5in}& -1/{2 N_{c} s_{14}} +1/{2 
s_{13}} 
  &\hspace{.5in}
  -1/{2 N_{c} s_{13}} +1/{2 s_{14}} \\            
     \hspace{.2in}{\rm (vi)}\hspace{.5in}& -1/{2 N_{c} s_{14}} 
+1/{2 
s_{13}} &\hspace{.5in} -1/{2 N_{c} s_{13}} +1/{2 s_{14}}  \\
\hspace{.2in}{\rm (vii)}\hspace{.5in} &-1/{2 N_{c} s_{14}} &\hspace{.5in} 
 1/{2 s_{14}}
\end{array}
\eqno (A.2)
$$

${\bf\mbox{\boldmath $\psi$}_{mn}^{\left(4,c\right)}}$: For 
each helicity 
channel,
$$ 
\begin{array}{lll}
\psi^{\left(4\right)}&=& |\tilde{c}|^{2} N_{c}^{2} \left[
 \left(A^{\left(0\right)}\right)^{2} + \left(B^{\left(0\right)}\right)^
{2} +
\frac{2}{N_{c}} A^{\left(0\right)} B^{\left(0\right)} \right] 
 \\ 
\psi_{12}^{\left(4,c\right)}&=& \psi^{\left(4\right)}/N_{c} 
- 
|\tilde{c}|^{2} N_{c} \left[ \left(A^{\left(0\right)}\right)^
{2} + 
\left(B^{\left(0\right)}\right)^{2} +
2 N_{c} A^{\left(0\right)} B^{\left(0\right)} \right] \nonumber 
\\
\psi_{13}^{\left(4,c\right)}&=& -\psi^{\left(4\right)}/N_{c} 
+
|\tilde{c}|^{2} N_{c} \left[ \left(A^{\left(0\right)}\right)^
{2} + 
N_{c} \left(B^{\left(0\right)}\right)^{2} \right] \nonumber 
\\ 
\psi_{14}^{\left(4,c\right)}&=& -\psi^{\left(4\right)}/N_{c} 
+
|\tilde{c}|^{2} N_{c} \left[N_{c} \left( A^{\left(0\right)}\right)^
{2} +
 \left(B^{\left(0\right)}\right)^{2} \right] \nonumber \\
\psi_{23}^{\left(4,c\right)}&=&\psi_{14}^{\left(4,c\right)},\,\,
\psi_{24}^{\left(4,c\right)}=\psi_{13}^{\left(4,c\right)}, \,\,
\psi_{34}^{\left(4,c\right)}=\psi_{12}^{\left(4,c\right)},\,

\end{array}
\eqno (A.3)
$$
where $A^{\left(0\right)}, B^{\left(0\right)}$ are
the sum of the corresponding contact and QCD terms as defined 
in (\ref{e101}).

${\bf\tilde{A}^{\left(1\right)},\tilde{B}^{
\left(1\right)}:}$ With the auxiliary functions,
$$
\begin{array}{lll}
F_{1}&=& 2 (N_{c}^{2}-1)/N_{c} \left[ 3 +
 \frac{3}{2} \ln \left(\frac{Q_{ES}^{2}}{
s_{14}}\right) + \frac{1}{2} \ln^{2} \left(\frac{Q_{ES}^{2}}{
s_{14}} \right) \right] \nonumber \\
F_{2}&=& 2  \left[ 4 + \frac{3}{2} \ln \left(
\frac{Q_{ES}^{2}}{
s_{13}}\right) + \frac{1}{2}\left( \ln^{2}
 \left(\frac{Q_{ES}^{2}}{
s_{13}} \right)-  \ln^{2} \left(\frac{Q_{ES}^{2}}{
-s_{12}} \right)\right) +\frac{3}{2} \ln \left(\frac{\mu^{2}}{
Q_{ES}^{2}}\right)  \right]  \nonumber \\
F_{3}&=&-F_{2}(s_{12} \leftrightarrow -s_{13}),
\end{array}
\eqno (A.4)
$$
and
$$
\begin{array}{lll}
F_{1o}&=&  -(N_{c}^{1}-1)/N_{c}^{2}  \left[ 4 + \frac{3}{2} 
\ln \left(
\frac{Q_{ES}^{2}}{
-s_{12}}\right) + \frac{1}{2}\left( \ln^{2} \left(\frac{Q_{ES}^
{2}}{
-s_{12}} \right)\right.\right. \nonumber \\
& &\left.\left.- \ln^{2} \left(\frac{Q_{ES}^{2}}{
s_{13}} \right)\right) +\frac{3}{2} \ln \left(\frac{\mu^{2}}{
Q_{ES}^{2}}\right)  \right] \nonumber \\
F_{2o}&=&  1/N_{c}  \left[5-N_{c}^{2} - \frac{3}{2} \ln \left(
\frac{Q_
{ES}^{2}}{
s_{14}}\right) + 3 \ln \left(\frac{Q_{ES}^{2}}{-s_{12}}\right) 
- 
\frac{1}{2}\ln^{2} \left(\frac{Q_{ES}^{2}}{
s_{14}} \right) \right. \nonumber \\
& &\left. +\ln^{2} \left(\frac{Q_{ES}^{2}}{-s_{12}} \right) 
-\left(1-
 \frac{N_{c}^{2}}{2}\right)\left( 
 \ln^{2} \left(\frac{Q_{ES}^{2}}{s_{13}} \right)  
-3  \ln^{2} \left(\frac{\mu^{2}}{Q_{ES}^{2}} \right)\right) 
 \right],
\nonumber 
\end{array}
\eqno (A.5)
$$
the contact terms are
$$
\begin{array}{crcl}
{\rm (i)} &\tilde{A}^{\left(1\right)}_{cont} &=&( 2 \eta_{0} 
- \eta_{1}/2) 
( F_{1}- F_{2}/N_{c}) + \eta_{1} F_{2}( p_{3}  \leftrightarrow
p_{4})  \\
& \tilde{B}^{\left(1\right)}_{cont} &=&( 2 \eta_{0} - \eta_{1}/2) 
F_{2} 
+ \eta_{1} ( F_{1}- F_{2}/N_{c})( p_{3}  \leftrightarrow p_{4}), 
 \\
{\rm (ii)}&\tilde{A}^{\left(1\right)}_{cont} &=&2( \eta_{0} 
+ \eta_{1}/4) 
\left[ F_{1}- F_{2}/N_{c} +
 F_{2}( p_{3}  \leftrightarrow p_{4}) \right]  \\
&\tilde{B}^{\left(1\right)}_{cont} &=& 2 ( \eta_{0} + \eta_{1}/4)\left[F_
{2} +  ( F_{1}- F_{2}/N_{c})
( p_{3}  \leftrightarrow p_{4}) \right],  \\
{\rm (iii)}&\tilde{A}^{\left(1\right)}_{cont} &=& \eta_{u} (
 F_{1} -F_
{3}/N_{c}) + \eta_{8u} ( F_{1o} -F_{2o}/N_{c})  \\
&\tilde{B}^{\left(1\right)}_{cont} &=&  \eta_{u} F_{3} + \eta_
{8u} F_{2o}, 
\nonumber \\
{\rm (iv)}&  \mbox{Replace} & &  \eta_{u} \rightarrow \eta_{d},\,\, 
\eta_
{8u} \rightarrow \eta_{8d}  \,\,\mbox{in}\,\, {\rm (iii)},  
\\
{\rm (v)}& \mbox{Replace} & &  (\eta_{0} + \eta_{1}/4)  \rightarrow 
\eta_
{uu} \,\,\mbox{in}\,\,{\rm (ii)}, \\
{\rm (vi)}&  \mbox{Replace}& &   (\eta_{0} + \eta_{1}/4) \rightarrow 
\eta_
{dd}
\,\,\mbox{in}\,\, {\rm (ii)},\\
{\rm (vii)}&\mbox{Replace}& &   ( 2 \eta_{0} - \eta_{1}/2) \rightarrow 
\eta_{ud}, \, \eta_{1} \rightarrow \eta_{ud}' \,\,\mbox{in}\,\,{\rm 
(i)}.
\end{array}
\eqno (A.6)
$$

{}For QCD, from the  one-loop amplitudes in \cite{r6},\newline 
(i),(vii);
$$
\begin{array}{lll}
\tilde{A}^{\left(1\right)}_{QCD}  &=& - \frac{1}{2 s_{14} N_
{c}} H_{1} 
\nonumber \\
\tilde{B}^{\left(1\right)}_{QCD}  &=& \frac{1}{2 s_{14}} ( H_
{1} +H_{2}),
\nonumber
\end{array}
\eqno (A.7)
$$
(ii),(v),(vi);
$$
\begin{array}{lll}
\tilde{A}^{\left(1\right)}_{QCD} &=& \frac{1}{s_{14}}\left[
 - H_{1}/N_{c} + \frac{s_{14}}{s_{13}}\left[\left(
H_{1}+ H_{2}\right) ( p_{3} \leftrightarrow p_{4})\right] \right]\nonumber 
\\
\tilde{B}^{\left(1\right)}_{QCD} &=& \frac{1}{s_{14}} \left[ 
H_{1}+ H_
{2}  - \frac{s_{14}}{ N s_{13}}
H_{1} ( p_{3} \leftrightarrow p_{4}) \right],
\nonumber
\end{array}
\eqno (A.8)
$$
(iii),(iv);
$$
\begin{array}{lll}
\tilde{A}^{\left(1\right)}_{QCD}  &=& - \frac{1}{2 s_{14} N_
{c}} K_{1} 
\nonumber \\
\tilde{B}^{\left(1\right)}_{QCD}  &=& \frac{1}{2 s_{14}} ( K_
{1} +K_{2}),
\nonumber
\end{array}
\eqno (A.9)
$$
where
$$
\begin{array}{lll}
H_{1}&=&  N_{c} (13/9 + \pi^{2}) - \frac{10}{9}  N_{f} +
\frac{1}{N_{c}}\left[ 8 + \frac{s_{14}}{2 s_{12}} ( 1+ s_{13}/s_
{12}) 
\left(
\ln^{2} \left( \frac{- s_{14}}{s_{12}}\right) + \pi^{2} \right) 
\right.
\nonumber \\
& & \left.+s_{14}/s_{12} \ln \left( \frac{s_{14}}{s_{13}} \right) 
\right] 
+ \ln\left(
\frac{Q_{ES}^{2}}{s_{14}}\right) \left[ - 3 N_{c} + \frac{11}{3} 
 N_{c} -2 \ln
\left( \frac{-s_{14}}{s_{12}}\right) - \frac{2}{3} N_{f} \right.\nonumber 
\\
& & \left. + 3/N_{c} +
2/N_{c} \ln\left( \frac{s_{12}}{-s_{13}} \right) \right]
- \ln^{2} \left( \frac{Q_{ES}^{2}}{s_{14}}\right)
 (  N_{c} - 1/N_{c})
+ 4 \pi \beta_{0} \ln\left( \frac{\mu^{2}}{Q_
{ES}^{2}}\right), \nonumber \\
H_{2}&=& -(N_{c}^{2}-1)/N_{c} \left[\frac{1}{2}
 \frac{s_{14}}{ s_{12}} ( 1+ s_{13}/s_{12}) \left(
\ln^{2} \left( \frac{s_{14}}{s_{13}}\right) + \pi^{2} \right) 
\right.
\nonumber \\
& & \left.+s_{14}/s_{12} \ln \left( \frac{s_{14}}{s_{13}} \right) 
+ 2 
\ln\left(
\frac{Q_{ES}^{2}}{s_{14}}\right) \ln \left( \frac{s_{12}}{-s_
{13}}\right)\right], 
\nonumber \\
K_{1}&=&  N_{c} (13/9 + \pi^{2}) - \frac{10}{9} N_{f} + 8/N_
{c} +
(N_{c}+1/N_{c}) \left[\frac{s_{14}}{2 s_{13}} ( 1+ s_{12}/s_
{13}) \right. 
\nonumber \\
& & \left.\left( \ln^{2} \left( \frac{s_{14}}{s_{13}}\right) 
+ \pi^{2} 
\right) 
 +s_{14}/s_{13} \ln \left( \frac{-s_{14}}{s_{12}} \right)\right] 
 + \ln\left(
\frac{Q_{ES}^{2}}{s_{14}}\right) \left[ - 3 N_{c} + \frac{11}{3} 
 N_{c} \right. 
\nonumber \\
& &\left. -2 N_{c} \ln \left( \frac{-s_{14}}{s_{12}}\right) 
- \frac{2}{3} 
N_{f}
 + 3/N_{c} +
2/N_{c} \ln\left( \frac{s_{12}}{-s_{13}} \right) \right] \nonumber 
\\
& &+ \ln^{2} \left( \frac{Q_{ES}^{2}}{s_{14}}\right) ( -  N_
{c} + 1/N_
{c})
+ 4 \pi \beta_{0} \ln\left( \frac{\mu^{2}}{Q_{ES}^{2}}\right), 
\nonumber 
\\
K_{2}&=& -(N_{c}^{2}-1)/N_{c} \left[\frac{s_{14}}{2 s_{13}} 
( 1+ s_{12}/s_
{13}) \left(
\ln^{2} \left( \frac{-s_{14}}{s_{12}}\right) + \pi^{2} \right) 
\right.
\nonumber \\
& & +\left. s_{14}/s_{13} \ln \left( \frac{-s_{14}}{s_{12}} 
\right) + 
2 \ln\left(
\frac{Q_{ES}^{2}}{s_{14}}\right) \ln \left( \frac{s_{12}}{-s_
{13}}\right)\right].
\nonumber
\end{array}
\eqno (A.10)
$$

\newpage
\parindent 0.0in
{\bf Figure Captions:}
\vspace{.5in}

\parindent 0.0in

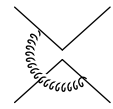
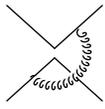
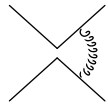
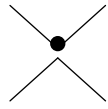
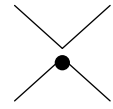
Fig. 1: One-loop diagrams from contact interactions. (g), (h)
are the counterterms for the octet current renormalization.

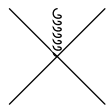
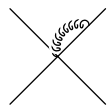
Fig. 2: One-loop diagram from dimension six operator. (b) is
suppressed by a factor $s/\Lambda^{2}$ compared to the 
 diagrams in Fig. 1.

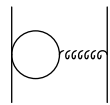
Fig. 3: Penguin diagrams induce  form factors in quark-gluon
vertex.

 
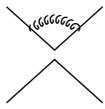
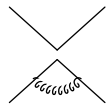
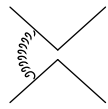
\end{document}